# Growth of Form in Thin Elastic Structures

Salem Al Mosleh[1], Ajay Gopinathan[2],* and Christian Santangelo[1][†]
*Department of Physics, University of Massachusetts Amherst, Amherst, MA 01003, USA[1] and*
*Department of Physics, University of California Merced, Merced, CA 95343, USA[2]*

Heterogeneous growth plays an important role in the shape and pattern formation of thin elastic structures ranging from the petals of blooming lilies to the cell walls of growing bacteria. Here we address the stability and regulation of such growth, which we modeled as a quasi-static time evolution of a metric, with fast elastic relaxation of the shape. We consider regulation via coupling of the growth law, defined by the time derivative of the target metric, to purely local properties of the shape, such as the local curvature and stress. For cylindrical shells, motivated by rod-like *E. coli*, we show that coupling to curvature alone is generically linearly unstable and that additionally coupling to stress can lead to stably elongating structures. Our approach can readily be extended to gain insights into the general classes of stable growth laws for different target geometries.

## I. INTRODUCTION

How physical processes establish the growth and form of biological structures was considered by D'Arcy Thomson almost a century ago [1]. Since then, there has been much progress explaining the different growth driven morphologies that appear in the natural world. These include understanding that the rippled edges of leaves [2], the ruffled petals of blooming lilies and other flowers [3, 4], and even the convolutions of the brain cortex may be driven by differences in growth rate between spatially distinct regions [5]. It is well known that heterogeneous insertion and deletion of material can lead to geometric frustration and shape change in synthetic tissues [6–8]. Yet one hundred years after D'Arcy Thomson's seminal work, there are still challenges and open problems. One such challenge is that of determining the connection between the dynamical growth law —where a tissue chooses to grow —and both the shape and stability of those tissues.

This raises the question of how growth laws are regulated in nature to ensure stable growth. Feedback is a commonly used mechanism in biology for ensuring stability, but it is not clear to what or how the growth laws need to be coupled, to ensure the robust growth of a stable structure. A particular example of this issue is the question of shape regulation in rod-like *E. coli*, which is still an open problem [9, 10]. Though the components of the molecular machinery responsible for cell wall growth and regulation have been identified [11], precisely how the nm-scale components within this network interact to produce a robust shape at the $\mu$m-scale is not completely understood. Feedback between cell wall insertion rate and curvature, which was shown to be present in *E. coli* [12], can in principle lead to stable cylindrical shapes. However, as demonstrated in Refs. [13] and [14], stress also affects cell wall insertion rate and can lead to growth which is different from what would be expected from a purely geometric coupling.

In this paper, we step back from the details of the growth process and consider a general framework for describing the growth of thin elastic structures that allows us to study stability. We assume that throughout the growth process, the material retains uniform thickness and Young's modulus. That is to say, it is still made of the same stuff, there is just more of it in some places and less in others. Mathematically, this growth process can be described as a change in the target metric of the shell or, alternatively, as the change in the local equilibrium lengths between points along the surface [7]. There are, of course, an infinity of ways that the target metric could change in time.

Here we consider regulating the growth by coupling the growth laws to purely local properties of the shape, such as the local curvature and stress. It is then possible to use considerations of symmetry and locality to make a curvature expansion and reduce the growth laws to only a few effective parameters. Our approach thus allows us to study the relationship between geometry and stress in determining the morphological stability of growing structures. Partially motivated by E. coli and partially for concreteness, we use our formalism to address the linear stability of elongating, cylindrical shapes as an example. Nevertheless, we develop principles that can be applied to morphology selection and stability in biological systems more generally.

This paper is organized as follows. In Sec. (II), we give a short overview of the required differential geometry. In Sec. (III) we consider the energetics of thin elastic shells, using the Helfrich Hamiltonian. In section (IV) we describe the growth process, and show how symmetry can help us organize the different possible growth laws. Sec. (V) studies the stability of elongating cylindrical shapes. After showing that purely geometric coupling alone is generically linearly unstable, we add the effect of coupling growth to stress and show that stability requires a combination of coupling to both curvature and stress. Finally, we conclude in Sec. (VI).

---
* e-mail: agopinathan@ucmerced.edu
† e-mail: csantang@physics.umass.edu

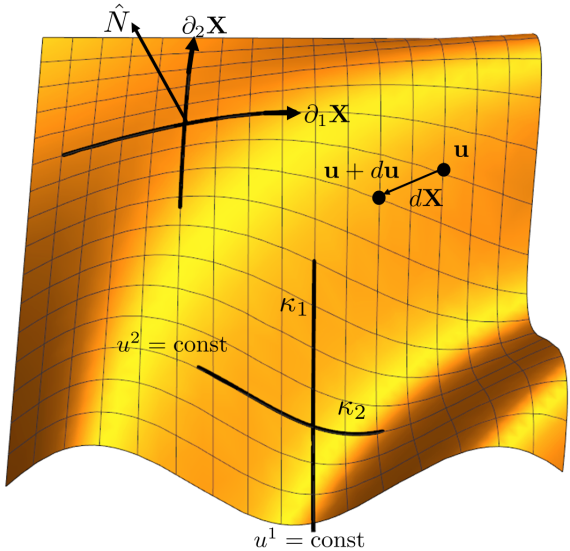

FIG. 1: (Color Online) $u^1$ and $u^2$ are the (arbitrary) coordinates chosen to parametrize the surface. Curves with constant coordinate values are shown. The vector $d\mathbf{X}$ is the displacement vector between the points parametrized by $\mathbf{u}$ and $\mathbf{u}+d\mathbf{u}$. The displacement vector satisfies $|d\mathbf{X}|^2 = d\ell^2$, which leads to the definition in Eq. (1). The unit vector $\hat{N}$ is normal to the surface at every point. We also show here the principal curvatures $\kappa_1$ and $\kappa_2$ defined roughly as the eigenvalues of the curvature tensor.

## II. DIFFERENTIAL GEOMETRY OF SURFACES

To establish notation, we give a brief overview of the differential geometry of surfaces in three dimensions [15, 16]. Throughout this paper we assume Einstein's summation convention, where repeated indices are summed unless otherwise stated. A surface embedded in 3D Euclidean space can be represented as a vector function of two variables, $\mathbf{X}(u^1, u^2) \equiv \mathbf{X}(\mathbf{u})$, as in Fig. (1). Information about the shape of the surface is encoded in the length and curvature of curves $\mathbf{u}(\ell)$ on the surface, parametrized by their arc length $\ell$. The length of any curve can be determined from the metric tensor through the relation

$$d\ell^2 = (\partial_i \mathbf{X} \cdot \partial_j \mathbf{X})\, du^i du^j \equiv g_{ij}\, du^i du^j \quad (1)$$

where $\partial_i$ is the partial derivative with respect to the coordinate $u^i$. Likewise the curvature tensor determines the curvature of curves in the direction normal to the surface through the relation (Fig. 1)

$$\kappa_N \equiv (\partial_i \partial_j \mathbf{X} \cdot \hat{N})\, \frac{du^i}{d\ell} \frac{du^j}{d\ell} \equiv b_{ij} \frac{du^i}{d\ell} \frac{du^j}{d\ell}. \quad (2)$$

Derivatives of tangent vectors can be expressed through the covariant derivative, formally defined on vectors as $D_i v_j = \partial_i v_j - \Gamma^k_{ij} v_k$ and $D_i v^j = \partial_i v^j + \Gamma^j_{ik} v^k$, where the Levi-Civita connection, $\Gamma^j_{ik}$ is given by

$$\Gamma^j_{ik} = \frac{1}{2} g^{jl} \left( \partial_i g_{kl} + \partial_k g_{il} - \partial_l g_{ik} \right). \quad (3)$$

On the other hand, the covariant derivative of a scalar function, $\phi(\mathbf{u})$, is the same as the coordinate derivative, so $D_i \phi = \partial_i \phi$. The failure of the covariant derivatives to commute measures the Ricci curvature of a surface, $R$. In particular,

$$[D_i, D_j] v^k = \frac{R}{2} \left( g_{jl} \delta^k_i - g_{il} \delta^k_j \right) v^l. \quad (4)$$

On the other hand, $D_i D_j \phi = D_j D_i \phi$.

We distinguish between the target metric $\bar{g}_{ij}$ and the actual metric $g_{ij}$. The target metric, $\bar{g}_{ij}$, encodes the local equilibrium lengths along any sufficiently small patch of the surface. Deviations of the actual metric from the target metric is encoded in the strain tensor $\epsilon_{ij} \equiv g_{ij} - \bar{g}_{ij}$. Growth can be represented as changes in the rest lengths on the surface. In other words, $\bar{g}_{ij}(t)$ will be time dependent. Here we do not consider time dependence of the target curvature tensor $\bar{b}_{ij}$.

Finally, we give the definition of the Gaussian, $K$, and mean, $H$, curvatures:

$$2H \equiv g^{ij}\, b_{ij} \quad \text{and} \quad K \equiv \frac{\det(b_{ij})}{\det(g_{ij})}, \quad (5)$$

where $g^{ij}$ is the matrix inverse of the metric $g_{ij}$, implying $g^{ij} g_{jk} = \delta^i_{\ k}$. Finally, note that by considering the matrix $b^i_{\ j} \equiv g^{ik} b_{kj}$, we can define two principle curvatures, $\kappa_1$ and $\kappa_2$, as the eigenvalues of $b^i_{\ j}$ along with their associated principle directions. These principle curvatures represent maximal and minimal normal curvatures of curves passing through a point and, thus, they are coordinate invariants. They are related to the Gaussian and mean curvatures through the relations $K = \kappa_1 \kappa_2$ and $2H = \kappa_1 + \kappa_2$.

It is well known that, if $g_{ij}$ and $b_{ij}$ satisfy compatibility conditions expressed through the Gauss-Codazzi-Mainardi equations, then they are sufficient to uniquely determine the surface up to rigid transformations. These compatibility relations specify that

$$\begin{aligned} R &= 2K \\ D_i b_{jk} &= D_j b_{ik}. \end{aligned} \quad (6)$$

In that sense, we have a complete characterization of any surface in three dimensions from $g_{ij}$ and $b_{ij}$ alone.

## III. ACCOUNTING FOR DYNAMICS

We start from the beginning (1666 AD), with Newton's laws [17] in a viscous medium

$$\sigma_M\, \partial_t^2 \mathbf{X}(\mathbf{u}, t) = -\gamma_D\, \partial_t \mathbf{X} - \frac{\delta_c E[\mathbf{X}]}{\delta_c \mathbf{X}} + \mathbf{f}(\mathbf{u}, t), \quad (7)$$

where $\sigma_M$ is the surface mass density and $\gamma_D$ is a drag coefficient, $E[\mathbf{X}]$ is the elastic energy, $\mathbf{f}$ is an externally applied force, and the functional derivative is defined as $\delta_c/\delta_c \mathbf{X} = \delta/(\sqrt{g}\, \delta \mathbf{X})$. With this definition $-\delta_c E[\mathbf{X}]/\delta_c \mathbf{X}$ gives the elastic force per unit area.

It is well known that the elastic energy of thin shells is composed of a stretching part, which is proportional to the thickness, $\tau$, and a bending part, which is proportional to $\tau^3$ [18–20]. Unlike stretching, bending deformations do not stretch the mid-surface of the shell. A quick experiment with paper will convince you that it costs much less energy to bend a thin sheet than it does to stretch it. Specifically, we take the elastic energy to be

$$E_{el} = \int d^2u \sqrt{g} \left[ \eta_S \; A^{ijk\ell} \; \epsilon_{ij}\epsilon_{k\ell} + \eta_B \; (H - H_0)^2 \right]. \quad (8)$$

We have defined $\eta_S \sim \tau$ and $\eta_B \sim \tau^3$ to absorb numerical factors. We also introduced the elasticity tensor $A^{ijk\ell} \equiv \lambda \bar{g}^{ij}\bar{g}^{k\ell} + 2\mu \bar{g}^{ik}\bar{g}^{j\ell}$, where $\lambda$ and $2\mu$ are the Lamé constants, which can be expressed in terms of Young's modulus $Y$ and Poisson's ratio $\nu$ as

$$\lambda \equiv \frac{Y\,\nu}{(1+\nu)(1-2\nu)}, \quad \text{and} \quad 2\mu \equiv \frac{Y}{1+\nu}. \quad (9)$$

The bending energy in Eq. (8) is slightly different from the standard choice $A^{ijk\ell}(b_{ij} - \bar{b}_{ij})(b_{k\ell} - \bar{b}_{k\ell})$. Since, for extremely thin shells, the bending energy is subdominant, we do not expect this choice to change the overall analysis. Furthermore, when the target curvature is isotropic, it can be written as $\bar{b}_{ij} = \bar{\kappa}\, \bar{g}_{ij}$, with $\bar{\kappa}$ being the two principle curvatures. With this choice the two energies become essentially equivalent as $\tau \to 0$.

In these expressions, the growth is implicit: $\bar{g}(t)$ is assumed to be a slowly-varying function of time. Due to the separation of growth and elastic time scales, we assume the elastic energy is minimized at each instant, with a quasi-static background metric. In the next section, we will account for the coupling between the target metric and the shape of the shell.

## IV. ACCOUNTING FOR GROWTH

In order to have a complete description of a growth process we need to specify how the background metric $\bar{g}_{ij}(t)$ changes with time. A generic class of growth laws can be described by giving the rate of change of the metric as a function of the shape, $\partial_t \bar{g}_{ij}(t) = F[\mathbf{X}]$. We will assume that $F[\mathbf{X}]$ is a local function of the shape, expressed in terms of the geometrical invariants already introduced. This is consistent with the notion that material insertion is determined from local information only.

There is of course, an infinite variety of possible growth laws consistent with this form; in this section we derive the most general growth law consistent with symmetries. Stated simply, locality is the assumption that the instantaneous change in the metric at a certain position depends only on quantities defined on the surface at that point. Coordinate invariance implies that the instantaneous change in the metric should be a rank-2 tensor on the surface. We assume that this tensor only depends on the local shape (principle of shape dependence) and an applied stress (strain) tensor, which severely restricts the form of the growth law. The constraints on the form of the growth law are coordinate invariance, locality and time homogeneity (absence of explicit time dependence).

We start by describing the geometry dependent terms in the growth law, then we turn to stress-coupled growth.

### A. Geometric Coupling

Deriving the geometric growth terms, in the vicinity of some arbitrary point with coordinates $\mathbf{u}$, is most conveniently done by transforming into a coordinate system where the metric at t = 0 is given by the identity matrix $\tilde{g}_{ij} = \delta{ij}$. This requirement however still does not fix the coordinate system. if the principle curvatures satisfy $\kappa_1 \neq \kappa_2$, then the coordinate axes are fixed by requiring the curvature tensor to have the form

$$\tilde{b}_{ij}(\mathbf{u}) = \begin{pmatrix} \kappa_1(\mathbf{u}) & 0 \\ 0 & \kappa_2(\mathbf{u}) \end{pmatrix}. \quad (10)$$

By locality, we mean that the mechanism responsible for generating the growth only has access to local shape information. To leading order in the vicinity of a point, the shape of the surface is defined by the two principle curvatures, and their directions. The principle directions can be taken without loss of generality to be in the $u^1$ and $u^2$ directions.

In an infinitesimal time step $dt$ the metric changes by an amount given by

$$\tilde{G}_{ij}(\kappa_1, \kappa_2) = \begin{pmatrix} f_1(\kappa_1,\kappa_2) & f_3(\kappa_1,\kappa_2) \\ f_3(\kappa_1,\kappa_2) & f_2(\kappa_1,\kappa_2) \end{pmatrix}, \quad (11)$$

so that the new metric is $\tilde{g}_{ij} = \delta_{ij} + dt\; \tilde{G}_{ij}$. As mentioned, in this coordinate system $\tilde{G}_{ij}$ can only be a function of $\kappa_1$ and $\kappa_2$ with no explicit time dependence due to the assumption of time homogeneity. To anticipate the form of this growth law in a general coordinate system we rewrite it in the form

$$\tilde{G}_{ij} = F_1(\kappa_1,\kappa_2)\,\delta_{ij} + a_0\, F_2(\kappa_1,\kappa_2)\,\tilde{b}_{ij} + F_3(\kappa_1,\kappa_2)\sigma^x_{ij}, \quad (12)$$

where $\sigma^x_{ij}$ is a Pauli matrix and $a_0$ is a length scale characterizing the size of the shell. When $\kappa_1 \neq \kappa_2$, the matrices $\delta_{ij}$, $\tilde{b}_{ij}$ and $\sigma^x_{ij}$ form a complete basis over the space of $2 \times 2$ symmetric matrices. In that case, it is possible to express a general choice of the functions $(f_1, f_2, f_3)$ in terms of $(F_1, F_2, F_3)$. However, in the case $\kappa_1 = \kappa_2$ the curvature tensor will also be proportional to the identity matrix. To avoid this problem, we might have replaced $\tilde{b}_{ij}$ with $\sigma^z_{ij}$ as a basis matrix. However, as we will show next, this is not ideal if the growth process depends purely on the local shape.

When $\kappa_1 = \kappa_2$, it is not possible to uniquely chose coordinate axes at that point because in that case, all directions are equivalent as far as local shape is concerned. Consequently, the growth process cannot favor any direction in this situation. Thus choosing $\tilde{b}_{ij}$ which is proportional to the identity matrix when $\kappa_1 = \kappa_2$ is the proper choice. In addition, since as $u^1 \to -u^1$, $F_3(\kappa_1, \kappa_2) \to -F_3(\kappa_1, \kappa_2)$, the term proportional $\sigma^x_{ij}$ is seen to violate chiral symmetry. Therefore, in the rest of this paper we will also take $F_3(\kappa_1, \kappa_2) = 0$. Finally, we write the growth law in a general coordinate system as

$$\partial_t \bar{g}_{ij} = F_1(H, K) \, g_{ij} + a_0 \, F_2(H, K) \, b_{ij}. \quad (13)$$

Note that we wrote $\kappa_1$ and $\kappa_2$ in terms of H and K.

We may simplify the growth law by assuming that there is a small length scale $\lambda$ controlling growth and sensing curvature. In the case of *E. coli*, this length scale is the nanometer scale of proteins as opposed to the $a_0 \sim \mu m$ scale of the bacteria. Compared to the length scale $\lambda$, the curvatures can be considered small, which motivates a curvature expansion of the functions

$$F_{(1,2)}(H, K) \approx \alpha_{(1,2)} + \beta_{(1,2)} \, \lambda \, (H - H_0) - \\ \gamma_{(1,2)} \lambda^2 \, (K - K_0) + \delta_{(1,2)} \lambda^2 \, (H - H_0)^2, \quad (14)$$

where we neglected terms of order $\lambda^3$. Note that terms of the form $|\kappa_1 - \kappa_2| = 2\sqrt{H^2 - K}$ are possible, but we neglect them due to their non-analyticity. For example, if $\mathbf{X}_S$ describes a sphere, then the rate of growth of a nearby surface $\mathbf{X}_S + \epsilon \, \delta \mathbf{X}$ will scale as $\partial_t g \sim O(\sqrt{\epsilon})$.

With that in mind, Eqs. (14) and (13) represents the most general geometrically-coupled growth law consistent with the assumed symmetries. Symmetry guarantees that spherical and cylindrical shapes will be fixed points of the evolution as we will show in Sec. V. However, instabilities may lead to spontaneous symmetry breaking and non-symmetric fixed points.

Next we will consider general growth laws in the presence of externally applied tensors, such as the strain (stress) tensor.

### B. Incorporating Stress Coupling

In this section, we seek growth laws that incorporate the role of the strain tensor, defined as $\epsilon_{ij} \equiv g_{ij} - \bar{g}_{ij}$.

We can write all the possible scalars and tensors that are consistent with our criteria. Raising and lowering are done only with $g_{ij}$ and $\epsilon \equiv g^{ij} \epsilon_{ij}$. The different scalars that we can construct are

$$H, \, \epsilon, \, b^{ij} \epsilon_{ij}, \, b_{ij} \, b^{ij}, \, b^i_k \, b^{ij} \epsilon_{jk}, \, \bar{\nabla}^i \bar{\nabla}^j \epsilon, \, \cdots \quad (15)$$

The tensors are

$$\epsilon_{ij}, \, g_{ij}, \, b_{ij}, \, \epsilon^k_i \, b_{kj}, \, b^k_i \, b_{kj}, \, \bar{\nabla}_i \bar{\nabla}_j \epsilon, \, \bar{\nabla}^k \bar{\nabla}_k \epsilon_{ij} \, \cdots \quad (16)$$

where $\nabla$ and $\bar{\nabla}$ are the covariant derivatives associated with the metrics $g_{ij}$ and $\bar{g}_{ij}$. Terms containing the covariant derivatives will be dropped since they are of order $O(\epsilon \, \lambda^2)$.

We can now construct the most general growth law neglecting terms of order $O(\lambda^3), O(\epsilon^2)$ and $O(\lambda^2 \, \epsilon)$. Concretely, we have

$$\partial_t \bar{g}_{ij} = \alpha_1 \, g_{ij} + \alpha_2 \, b_{ij} + \beta_1 \, H \, g_{ij} + \beta_2 \, H \, b_{ij} \\ -\gamma_1 \, K \, g_{ij} - \gamma_2 \, K \, b_{ij} + \sigma_1 \, \epsilon_{ij} + \sigma_2 \, \epsilon \, g_{ij} + \sigma_3 \, H\epsilon_{ij} + \\ \sigma_4 \, H \, \epsilon \, g_{ij} + \sigma_5 \, \epsilon \, b_{ij} + \sigma_6 \, b^{k\ell} \epsilon_{k\ell} \, g_{ij} + \sigma_7 \, \epsilon^k_i \, b_{kj}. \quad (17)$$

In appendix (A) we will try to gain intuition of the various terms in this growth law.

## V. LINEAR STABILITY OF ELONGATING CYLINDERS

We now analyze the linear stability of elongating cylindrical shells under the growth law in Eq. (17). However, this analysis could be applied more generally, for example, to spherical and leaf-like shells. As mentioned earlier, spherical, planar, or cylindrical symmetries will tend be preserved under time evolution. Therefore surfaces with these symmetries form a kind of generalized fixed point. We say generalized because they may still be evolving, as in the case of the elongating cylinder. However, instabilities may cause spontaneous symmetry breaking to non-symmetrical shapes. Using linear stability analysis we can determine the parameter values for which a given symmetry is linearly stable.

We will start our analysis by considering purely geometric coupling, later we will consider the effect of adding the stress coupling terms.

### A. Purely Geometric Coupling

As mentioned in Sec. IV A, we will ignore terms which have dimensions of inverse length, since these terms will be multiplied by a small length scale $\lambda$. The growth law in the purely geometric case is

$$\partial_t \bar{g}_{ij} = \alpha_1 \, g_{ij} + \alpha_2 \, a_0 \, b_{ij} + \beta_1 \, a_0 \, (H - H_0) \, g_{ij} + \\ a_0^2 \, \beta_2 \, (H - H_0) \, b_{ij} - a_0^2 \, \gamma_1 \, (K - K_0) \, g_{ij}. \quad (18)$$

Here $H_0$ and $K_0$ represent some time-independent target curvatures, and so their definition can be absorbed into $\alpha_{(1,2)}$. We will take them to be the mean and Gaussian curvatures of the fixed point solution. For a sphere with initial radius $a_0$, we will have $H_0^2 = K_0 = 1/a_0^2$. Naturally for a flat fixed point we would chose $K_0 = H_0 = 0$. Finally, for an elongating cylinder with radius $a_0$ we will chose $K_0 = 0$ and $H_0 = -1/2a_0$. We've neglected a term proportional to $(H - H_0)^2$ since it will not be important for linear stability.

We start by seeking elongating cylindrical solutions

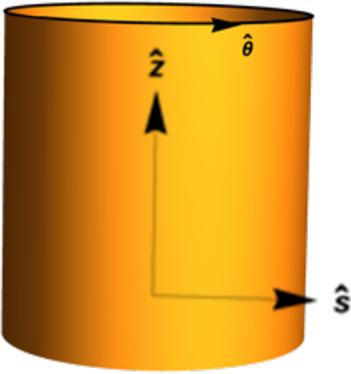

FIG. 2: A coordinate system describing an elongating cylinder. The coordinate z, along the $\hat{z}$ direction, is normalized such that $z \in (0, 1)$.

of the form

$$\mathbf{X}_{Cyl} = a\, \hat{s}(\theta) + z\, L(t)\, \hat{z}$$
$$\bar{g}_{ij} = \begin{pmatrix} L_0^2(t) & 0 \\ 0 & a_0^2 \end{pmatrix}, \quad (19)$$

where $\hat{s}$, $\hat{z}$ and $\hat{\theta}$ are the cylindrical basis vectors (see Fig. 2). Here $z \in [0, 1]$ and $L_0(t)$ is the time dependent length of the shell. The definition of $z$ is convenient because it allows us to consider a boundary value problem on the domain $z \in [0, 1]$ instead of a time dependent domain $[0, L_0(t)]$. Furthermore, note that $z$ and $\theta$ are thought of as material coordinates. In other words, if a given point on the shell was properly tagged and followed, its trajectory would be given by $\mathbf{X}(z, \theta, t)$, where $(z, \theta)$ are the coordinates on the initial cylinder at time $t = 0$.

In order to study the stability of the solution in Eq. (19), we need to first verify that it is a solution of the growth and elastic equilibrium equations in the absence of stress coupling. We assume $\bar{H} = H_0$, which implies that bending and stretching energies prefer the same radius $a_0$. By plugging Eq. (19) into the elastic energy and minimizing with respect to $a$ and $L$ we find, as expected, $a = a_0$ and $L = L_0$. Of course, if $\bar{H} \neq H_0$ or we had nonzero pressure $p \neq 0$, then there will be small corrections to this answer. To find the time dependence of the length, we plug the ansatz Eq. (19) into the geometric growth law Eq. (18). Concretely, we get the conditions

$$\alpha_2 = \alpha_1, \quad \alpha_1 = 2\, R, \quad \text{and } L_0(t) = \ell_0\, e^{R_0\, t}. \quad (20)$$

The first condition results from the requirement of fixed radius. As explained in appendix (A), a fixed radius emerges due to a balancing between isotropic expansion and inward volume contracting growth terms. A more physical way to say this, is that the $\alpha_2$ term, which is proportional to $b_{ij}$, represent a slowing down of material insertion along any curved direction. Unlike the cylindrical case, a sphere has two nonzero curved directions, which implies that this balancing would lead to a halting of growth in all direction.

At this point we introduce a perturbation to the elongating cylinder, which has the form

$$\mathbf{X}(z, \phi, t) = a\left[1 + \rho(z, \phi, t)\right]\, \hat{s} +$$
$$L(t)\left[z + h(z, \phi, t)\right] \hat{z} + a\, \psi(z, \phi, t) \hat{\phi},$$
$$\bar{g}_{ij} = \begin{pmatrix} L_0(t)^2 + G_{zz}(z, \phi, t) & G_{z\phi}(z, \phi, t) \\ G_{z\phi}(z, \phi, t) & a^2 + G_{\phi\phi}(z, \phi, t) \end{pmatrix} (21)$$

Plugging this into the growth equations (18) and assuming the conditions (20) gives us a set of three coupled partial differential equations, which are second order in spacial coordinates and first order in time. However, since we have six unknown function in Eq. (21), we need to use the three elastic equilibrium equations. This is more easily done in Fourier space, which is possible since the equations are linear and the target solution has cylindrical symmetry. Specifically, we have

$$\rho(z, \phi, t) = \sum_m \int \frac{dq}{4\pi^2}\, \rho_{mq}(t)\, e^{im\phi}\, e^{iqz} \quad (22)$$

And similarly for other functions. Periodic boundary conditions in the $\phi$ direction are implied in this expansion. A more realistic basis in the z direction would be $\sin(n\pi z)$, wigth $n = 1, 2, \cdots$. However, the expression for the growth rate of perturbations will not depend on this choice as long as we keep in mind that $q_{min} \sim 1/\pi$.

We will solve the equations in two steps. First, we solve the elastic equilibrium equations for the components of $\bar{g}_{ij}$, then we use the growth equations to find the growth rate of radial perturbations $R_\rho(m, q) \equiv \dot{\rho}_{mq}/\rho_{mq}$. The elongating cylinder will be stable if $R_\rho < 0$ for all excitable modes. Since the resulting algebra is too long to show here, we will only show the results in the vanishing thickness ($\eta_B \to 0$) limit. However the finite thickness results will be plotted and discussed.

We first need to find the elastic equilibrium equations. To leading order, the elastic energy can be written as

$$E_{el} = \sum_{\mathbf{m}} \int \frac{dz\, d\phi\, d^2\mathbf{q}}{4\pi^2} e^{i(m_1-m_2)\phi}\, e^{i(q_1-q_2)z}\, E_{\mathbf{mq}}[\rho_{\mathbf{mq}}, h_{\mathbf{mq}}, \psi_{\mathbf{mq}}] = \sum_m \int dq\, E_{mq}[\rho_{mq}, h_{mq}, \psi_{mq}], \quad (23)$$

where $\mathbf{m} = \{m_1, m_2\}$ and $\mathbf{q} = \{q_1, q_2\}$. We can find the equilibrium equations by taking the derivatives of the energy with respect to the independent variable. Specif-

ically

$$\frac{\delta E_{mq}}{\rho_{mq}} = \frac{\delta E_{mq}}{h_{mq}} = \frac{\delta E_{mq}}{\psi_{mq}}. \tag{24}$$

for vanishing thickness ($\tau, \eta_B \to 0$), the solution to Eq. (24) is given by

$$\rho_{mq} = \frac{m^2\,G_{zz} - 2\,m\,q\,G_{z\phi} + q^2\,G_{\phi\phi}}{2\,a_0\,q^2},$$

$$h_{qm} = -\frac{i\,G_{zz}}{2\,q\,L_0^2} \quad \text{and}$$

$$\psi_{mq} = i\frac{m\,G_{zz} - 2\,q\,G_{z\phi}}{2\,a_0^2\,q^2}, \tag{25}$$

which is an isometry of the metric given in Eq. (21). Alternatively, we can invert Eq. (25) to eliminate the components of $\bar{g}_{ij}$ from the growth equations. After plugging the resulting answer in the growth law, we obtain three first order ODEs for the functions $\rho_{mq}(t)$, $h_{mq}(t)$ and $\psi_{mq}(t)$. For the case $\eta_B \to 0$, the growth equations become

$$\frac{\dot{\rho}_{mq}}{\rho_{mq}} = -\frac{1}{4}\left(\Gamma_1 + q_P^2\,\Gamma_2 + m^2\,\Gamma_3 + \frac{m^2(m^2-1)\Gamma_4}{q_P^2}\right),$$

$$\dot{h}_{mq}(t) = -\frac{im}{4}\left(\Gamma_5 + \frac{(m^2-1)\,\Gamma_4}{q_P^2}\right)\rho_{mq}(t) \quad \text{and}$$

$$\dot{\psi}_{mq}(t) = -\frac{im}{4}\left(\Gamma_5 + \frac{(m^2-1)\,\Gamma_4}{q_P^2}\right)\rho_{mq}(t), \tag{26}$$

were we introduced the physical wavenumber as $q_P \equiv a_0\,q/L_0(t)$. With this definition, the instantaneous wavelength of the deformation is $\lambda_P = a_0(2\pi/q_P)$. It is interesting to note that even though $q$ is time independent, $q_P$ is not. This is due to the stretching of the wavelengths during elongation. We have also introduced the $(q_P, m)$ independent rates $\Gamma_i$, which are given by

$$\Gamma_1 = \beta_2 - \beta_1 - 4R_0, \quad \Gamma_2 = 2\gamma_1 + \beta_1 - \beta_2,$$
$$\Gamma_3 = 2\,\gamma_1 + 2\,\beta_1 - \beta_2, \Gamma_4 = \beta_1$$
$$\text{and} \quad \Gamma_5 = 2\gamma_1 + \beta_1 - 4R_0. \tag{27}$$

Note that $h_{mq}$ and $\psi_{mq}$ satisfy the same equation and they both approach a constant value when $\rho_{mq} \to 0$. It turns out, this last fact is true for any coordinate invariant growth law. This is easy to see by considering Eq. (21) with $\rho, \psi, \dot{h} \to 0$, for which the surface becomes $\mathbf{X} = a_0\,\hat{s} + L(t)(z + h_0(z,\phi))$. It's not too hard to convince yourself that this surface is still cylindrical. In other words, this deformation is equivalent to a coordinate transformation. Since the growth law is coordinate independent, a deformation with arbitrary $h_0(z,\phi)$ is a fixed point solution, which, just as Eq. (19), describes an elongating cylinder. This explains why $\dot{h}_{mq} = 0$ when $\rho_{mq} = 0$. A similar argument can be given for $\psi_{mq}$ in the linear regime.

Thus, a sufficient and necessary condition for linear stability is $\rho_{mq}(t \to 0) = 0$, or that $R_\rho \equiv \dot{\rho}_{mq}/\rho_{mq} < 0$ for all permissible wavenumbers $q_P$ and $m$. However, there is a subtlety associated with this stability condition.

As mentioned above, $q_P$ is time dependent and as the wavelength of a solution is stretched, its rate of growth will also change. In particular, it is conceivable that $R_\rho > 0$ for a given $q_P$, but the solution is still stable. This is because the solution only experiences this instability for a short period of time before the physical wavelength $q_P$ changes to value where $R_\rho < 0$. Keep in mind however, that this only happens if $R_\rho > 0$ for a small range of $q_P$, at the onset of an instability. In addition, some long wavelength instabilities might not be realized until the shell length $L_0(t)$ reaches a certain value.

Keeping all of that in mind, we derive the necessary condition for stability of all modes as $t \to \infty$, which implies $R_\rho < 0$ for all $q_P$ and $m$ up to a high cutoff.

From Eq. (26) we find the rate of growth to be

$$R_\rho = -\frac{1}{4}\left(\Gamma_1 + q_P^2\,\Gamma_2 + m^2\,\Gamma_3 + \frac{m^2(m^2-1)\Gamma_4}{q_P^2}\right). \tag{28}$$

A more complicated expression also exists in the finite thickness regime which, interestingly, also depends on the combinations $\Gamma_i$. Apart from the growth law parameters, $R_\rho$ depends only on the physical wavelengths in units of $a_0$. The wavenumber $m$ can in principle be any integer, however there will be a high cutoff value corresponding to a small length scale. $q_P$ on the other hand will have a lower bound as well, corresponding to the finite size $L_0(t)$. Interestingly, this lower bound is time dependent, decreasing with time. To zeroth order, we will require stability for all $m$ and all $q_P$ without bounds.

We first find the stability region in parameter space for the zero thickness case. Then we will see how finite thickness changes the situation. It is clear from Eq. (28) that we must have $\Gamma_1, \Gamma_2, \Gamma_3, \Gamma_4 > 0$. This leads to the conditions

$$\beta_1 > 0, \quad \beta_2 > 4R_0 + \beta_1, \quad 2\gamma_1 > \beta_2 - \beta_1 \tag{29}$$

This defines three planes that bound the stability region in parameter space. Fig. (3) shows a cross section of this region along with modes of instability when $\sigma_1, \sigma_2, \eta_B > 0$, which will be discussed shortly. In fact, this is also the region of stability for the finite thickness and stress coupling cases.

Now we turn to the interesting question of what happens near the three boundary surfaces of the stability region. Consider approaching the boundary $\Gamma_1 = 0$, while $\Gamma_2, \Gamma_3, \Gamma_4 > 0$. It is easy in this case, to see that the rate is maximized when $m = 0$ and $q_P \to 0$. This is illustrated in Fig. (4). It can also be seen readily from Eq. (28) that crossing the boundary $\beta_1 = 0$ results in modes with high $m \sim m_{cutoff}$ and small $q_P \sim a_0\pi/L_0$ dominating the shape. Here $m_{cutoff}$ is the mode number at which our long wavelength approximation fails. However, finite thickness regularizes this behavior. As can be seen from Fig. (5) the instability in the finite thickness case happens at $m = 2$ and $q_P \sim 0.2$.

We may also get an instability that favors modes of high $q_P \sim q_{cutoff}$ by setting $\Gamma_2 < 0$. Unfortunately in this case, the bending energy does not regularize the behavior at large q, and in fact, seems to make these modes more unstable. In particular, in the finite thickness case we have

$$\lim_{q_P \to \infty} R_{\rho B} = -\lim_{q_P \to 0} R_{\rho B} = -\lim_{m \to \infty} R_{\rho B} = 4\,R, \quad (30)$$

where the index B in $R_{\rho B}$ is added to emphasize that bending energy is considered. We see from Eq. (30) that the effect of bending energy is to make the high $m$ modes always stable, while high $q_P$ modes are unstable for all parameter values (see Fig. 6). This non-intuitive result is one of the main contributions of this paper.

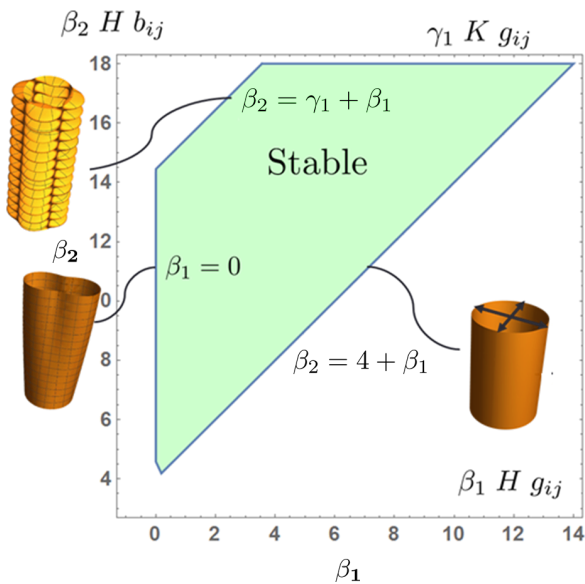

FIG. 3: This figure shows the stability region in the $(\beta_1, \beta_2)$ plane with $\gamma_1 = 15$ and stress coupling $\sigma_1 = \sigma_2 = 10$. Here and in all plots $R_0 = 1$, $a_0 = 1$, $\nu = 1/3$, $\eta_B = 0.01^3$ and $\eta_S = 0.01$. We also show the nature of the instabilities when crossing the different boundaries. The nature of these instabilities depends on $\sigma_1, \sigma_2, \eta_B > 0$, however the region itself would look the same in the case $\sigma_1, \sigma_2, \eta_B = 0$.

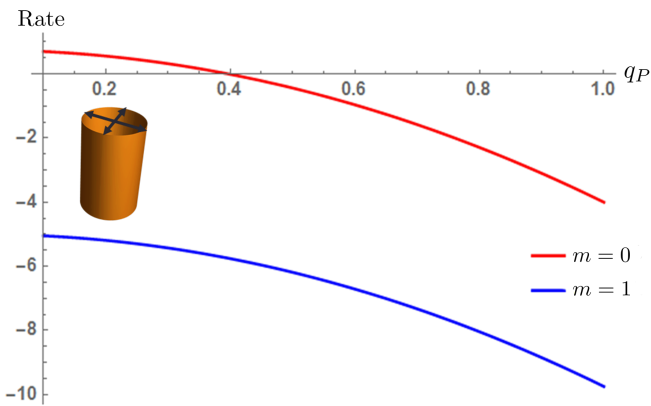

FIG. 4: This figure shows the growth rate as function of $q_P$ for different values of $m$ when $\beta_2 < \beta_1 + 4R_0$. Note that the maximum rate happens at $m = 0$ and $q \sim 0$.

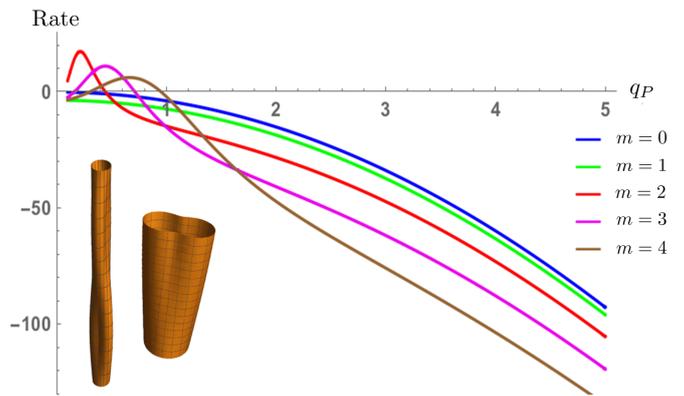

FIG. 5: This figure shows the growth rate as function of $q_P$ for different values of $m$ when $\beta_1 < 0$. Note that the maximum rate happens at $m = 2$.

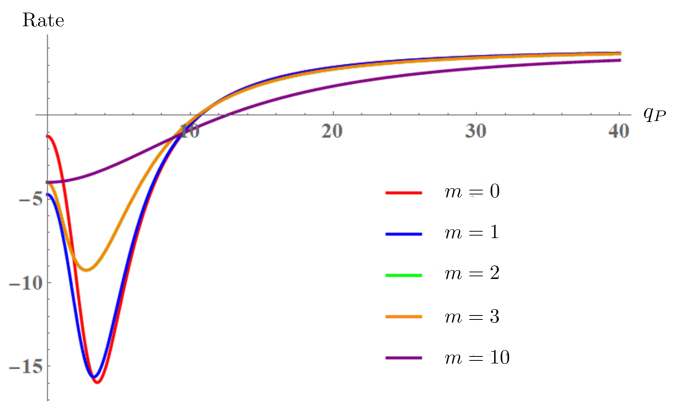

FIG. 6: This figure shows that all modes are unstable in the absence of stress coupling for high values of $q_P > a_0/\tau$. Here we set $\eta_S^3 = \eta_B = 0.3^3$, $\beta_1 = 5$, $\beta_2 = 14$, $\gamma_1 = 9$, $\sigma_{1,2} = 0$ and $\nu = 1/3$. Fig. 8 shows how stress coupling modifies and stabilizes this behavior.

The reason this is counter intuitive is because bending is expected to suppress modes of small wavelength rather than enhance them, which is true in the static setting. In a growing shell, to suppress the small wavelength fluctuations, their growth in the target metric must be suppressed. In the absence of bending, this suppression happens through the $\Gamma_2$ term in Eq. (28). As we show in appendix (C), bending suppresses this term indirectly by inhibiting high curvature modes. Amazingly, by suppressing high curvature modes, you allow them to grow further in the target metric.

Regardless of the source of these instabilities, a growing shell – such as E. coli – must find a way to avoid these instabilities. One possibility is that the small wavelength cutoff, $\lambda$, is on the order of the thickness of the shell. This is a reasonable possibility since the expansion of the energy in powers of thickness breaks down. For wavelengths

that are close to the thickness, the rate behaves as

$$R_{\rho B}(q_P \to \frac{a_0}{\tau}) = -4 \; \frac{\Gamma_2 - R_0 \; (1-\nu^2)}{1-\nu^2}. \quad (31)$$

Therefore, in the absence of stress coupling we must require that $\Gamma_2 > R_0 \; (1-\nu^2)$ and $q_P \lesssim \tau$ to achieve stability. The appearance of the material parameter $\nu$ (Poisson's ratio) in this expression is due to its effect on the response of the shape to the bending force, which in turn affects the growth rate.

Another, more robust way to stabilize small wavelength fluctuations is accomplished by coupling stress to the growth, which we turn to next.

### B. Stress Coupling To The Rescue

As we have seen in appendix (A), the term $\sigma_1 \; g_{ij}$ with $\sigma_1 > 0$ tends to make the target metric grow to comply with the applied force. So if the applied force is bending, then we may expect that adding stress coupling can lead to suppression of the modes $q_P \gtrsim \tau$. After ignoring terms of order $O(\lambda \; \epsilon)$ as described before, we step through the calculation in a similar manner to that described above. We eventually get

$$\lim_{q \to \infty} R_{\rho B} = 4 \; R_0 - \left( \sigma_1 + \frac{1-2\nu}{1-\nu} \sigma_2 \right). \quad (32)$$

In other words if the stress coupling is strong enough, then small wavelength modes are always stable no matter what parameters you use. It can also be shown, for small enough thickness, that the stability region in this case is the same as before (Figs. 3 and 7). In addition, the $\beta_1 < 0$ and $\beta_2 < \beta + 4 \; R$ instabilities still look the same (see Figs. 4 and 5). However, as Fig. (8) shows, when we cross the $\Gamma_2 = R_0 \; (1-\nu^2)$ plane, the instability will not start at the highest $q_P$ modes as before. In fact, it will happen typically for $m = 2$ and $q_P \sim O(10)$.

Thus, stress coupling enhances the stability against small wavelength deformations. And so, having both geometric and stress couplings can lead to stability of an elongating cylinder against all modes for a broad range of parameters.

Finally, we mention the possibility of stability with purely stress coupling. In other words, growth would stop in the absence of stress. In this case, elongation can be accomplished either due to pressure or incompatibility between the target metric and curvature tensors. However, we will not consider these possibilities in detail here since they will be part of future work. For now, we will consider the simpler question of what happens when we set the couplings $\beta_1 = \beta_2 = \gamma_1$. In this case, it can be shown that $R_\rho(m=0) = R_0 + O(\tau^2)$, which is unstable for all $\sigma_1, \sigma_2$. Roughly speaking, this indicates that geometric coupling might be important for the stability of long wavelength modes.

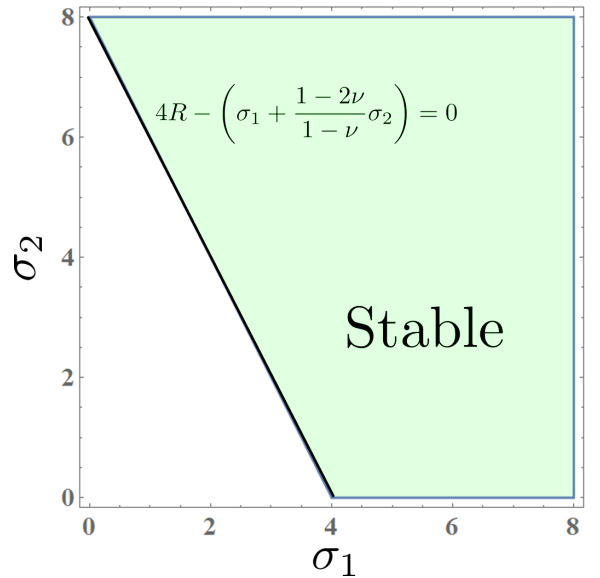

FIG. 7: This figure shows the stability region in the $(\sigma_1, \sigma_2)$ plane with $\gamma_1 = 15$, $\beta_1 = 6$ and $\beta_2 = 16$. Here and in all plots $R_0 = 1$, $a_0 = 1$, $\nu = 1/3$ , $\eta_B = 0.01^3$ and $\eta_S = 0.01$.

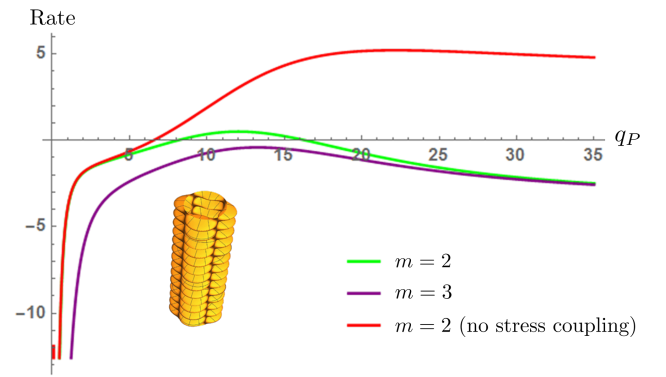

FIG. 8: This figure shows the growth rate as function of $q_P$ for different values of $m$ when $\Gamma_2 < R_0 \; (1-\nu^2)$ and $\sigma_1, \sigma_2 > 4R$. Note that the maximum rate happens at $m=2$ and at a finite value of $q_P \sim O(10)$. For comparison, we have also included a plot of the rate in the absence of stress coupling.

### VI. CONCLUSION

Growing elastic shells appear in a wide variety of contexts ranging from synthetic and natural shape changing materials that can be activated by spatially controlled swelling [7, 8] all the way to the growth of biomaterial sheets and planar tissues by the addition of material and proliferation of cells respectively [3, 4, 10]. In this paper, we have addressed for the first time, the dynamics of such growing sheets and the consequences for their stability. We have assumed that a growing shell, to which material is being added and removed, can be described with a slowly changing target metric. This is because, as a shell's structure rearranges, the natural distances be-tween points in the shell change. Within this setup there are infinitely many ways that the metric can change with time. It can change in a prescribed shape-independent way, as done in experiments like [7, 8], it can be coupled to an externally applied field like a stress tensor or an ordered template [13] or it can be dependent purely on shape as in Eq. (13).

Regulation of such growth to yield a desired structure typically requires a control mechanism. Such control mechanisms could couple the processes driving the growth to global properties of the shape, or to local properties of the shape, allowing the material to act locally and think globally. Therefore, in this paper we explored the coupling of the change in the metric to local properties of the sheet - the local shape, and a stress tensor. Symmetry and locality arguments help reduce the space of possible metric changes down to the form given in Eq. (17).

After constructing this general growth law, as a first step, we analyzed the linear stability of an elongating cylinder under purely geometric coupling (Sec. V A). Surprisingly, we found that for any choice of model parameters, modes with wavelengths on the order of the thickness ($q_P \sim 1/\tau$) cannot be stabilized (see Figs. 6,7). This unexpected result means that a growth law that is only shape-dependent cannot lead to an elongating cylinder that is linearly stable to small wavelength fluctuations.

Since biological systems appear to be able to solve this problem, we consider two possibilities. First, there might be a cutoff beyond which the assumptions under which our growth law is derived will not be valid. One could imagine, for example, that nonlinearities might result in the suppression of instabilities. However, even in such cases, one might expect to see the vestiges of the onset of the instability. This raises the intriguing possibility that such arrested instabilities could be used to create small scale patterns. To more robustly stabilize the growth, we consider a second mechanism, stress coupling, discussed in Sec. (V B). In particular when the effects of stress coupling are included we find that these small wavelength modes become universally stable, as shown in Figs. (8 and 7). This is because the stress coupling terms tend to make the target metric grow in a viscoelastic-like way to conform with the applied forces as discussed in Appendix A. In this situation, the applied forces are the bending forces (in E. coli, turgor pressure contributes as well), and since small wavelength modes contribute a lot of bending energy they will be suppressed. Note that both the applied force and the stress coupling contribute to this result.

Interestingly, it was shown in [14] that coupling to areal strain alone can result in straightening of a bent rod. However, under this growth law a shell might still be unstable with respect to different modes of deformation. Experiments involving controlled perturbations of the growth laws can yield a significant amount of information on the exact nature of the couplings. To this end, we are currently working on fitting the parameters of the model to experiments where bacteria are subjected to bending forces and oscillatory osmotic shocks resulting in perturbations in localizaation and dynamics of growth [13, 21]. One could also imagine directly probing the instabilties by growing E. coli in confining geometries with shapes of a specific wavelength in the z and $\phi$ directions. The exact form of $R_\rho$ could then be compared to the results of the experiment.

Finally, it is also conceivable that a certain shape cannot be stabilized at all, just as we've seen that, with purely geometric coupling, an elongating cylindrical shell would always be unstable to small wavelength fluctuations. While flat, cylindrical and spherical shapes are fixed points of the growth law due to symmetry, an interesting project would be a characterization of all the possible stable shapes within this framework and relating them to the kinds of patterns observed in nature.

We acknowledge useful and enlightening conversations with B. Chen, L. Mahadevan, A. Amir and K.C. Huang. CS and SA were funded by the National Science Foundation under award DMR-1507377 and AG was partially supported by National Science Foundation NSF grant DMS-1616926, a James S. McDonnell Foundation Award and in part by the NSF-CREST: Center for Cellular and Bio-molecular Machines at UC Merced (NSF-HRD-1547848).

---

[1] D'Arcy T. *On growth and form*. Cambridge, University Press, New York, Macmillan. (1945).

[2] M. Marder, E. Sharon, S. Smith and B. Roman. *Theory of edges of leaves.* **62**, 4. (2014)

[3] H. Liang, L. Mahadevan. *Growth, geometry, and mechanics of a blooming lily*, Proc. Natl. Acad. Sci. USA **108**, 5516-5521 (2011).

[4] Dervaux, J. and Amar, M.B., "Morphogenesis of growing soft tissues," Phys. Rev. Lett. **101**, 068101 (2008).

[5] Tallinen T, Chung J., Rousseau F, Girard N, Lefvre J and Mahadevan L. *On the growth and form of cortical convolutions.* **12**, 588-593. (2016).

[6] Castle, T., Cho, Y., Gong, X., Jung, E., Sussman, D.M., Yang, S. and Kamien, R.D., 2014. Making the cut: Lattice kirigami rules. Physical review letters, 113(24), p.245502.

[7] Y. Klein, E. Efrati and E. Sharon. *Shaping of elastic sheets by prescription of non-Euclidean metrics*. Science **315**, 1116-1120 (2007).

[8] Kim J, Hanna JA, Byun M, Santangelo CD, Hayward RC. "Designing responsive buckled surfaces by halftone gel lithography," Science **335**, 1201-1205. (2012).

[9] Amir A., Teeffelen S. *Getting into shape: How do rod-like bacteria control their geometry?* Syst Synth Biol. **8(3)**,

[10] Chang F, Huang KC. *How and why cells grow as rods.* BMC Biol. **12**, 54. (2014).
[11] D. Scheffers and Mariana G. Pinho. *Bacterial Cell Wall Synthesis: New Insights from Localization Studies.* **69(4)**, 585-607. (2005).
[12] T. Ursella, J. Nguyenb, R. D. Mondsa, A. Colavinc, G. Billingsd, N. Ouzounove, Z. Gitaie, J. W. Shaevitzb and K. C. Huang. *Rod-like bacterial shape is maintained by feedback between cell curvature and cytoskeletal localization.* PNAS. **111(11)**, E1025-E1034. (2014).
[13] Amir A, Babaeipour F, McIntosh D, Nelson D, and Jun S. *Bending forces plastically deform growing bacterial cell walls.* PNAS. **111**, 57785783. (2014).
[14] Wong F, Renner LD, zbaykal G, Paulose J, Weibel DB, van Teeffelen S, Amir A. *Mechanical strain sensing implicated in cell shape recovery in Escherichia coli*, Nature Microbiology. **2**, 17115 (2017)
[15] M.P. Do Carmo, *Differential Geometry of Curves and Surfaces* (Prentice-Hall, 1976).
[16] R.M. Wald, *General Relativity* (University of Chicago Press, 1984).
[17] Newton, Isaac, 1642-1727. Newton's Principia: The Mathematical Principles of Natural Philosophy. New-York :Daniel Adee, 1846.
[18] Augustus Love. *A treatise on the mathematical theory of elasticity.* 1, (1892). <hal-01307751>
[19] P. Audoly and Y. Pomeau, Elasticity and Geometry (Oxford University Press, Oxford, 2010).
[20] L.D. Landau and E.M. Lifshitz, Theory of Elasticity (Butterworth-Heinemann, Oxford, 1986).
[21] ER Rojas, JA Theriot, KC Huang, *Response of Escherichia coli growth rate to osmotic shock,* PNAS **111** 7807-7812 (2014).
[22] D. A. Quint, A. Gopinathan, and Gregory M. Grason. *Shape Selection of Surface-Bound Helical Filaments: Biopolymers on Curved Membranes.* Biophys. J. **111(7)** 1575-1585. (2016).
[23] Wang, Siyuan et al. *Cell Shape Can Mediate the Spatial Organization of the Bacterial Cytoskeleton.* Biophys. J. **104(3)**, 541-552 (2012)
[24] Y. Fily, A. Baskaran, M. F. Hagan. Active Particles on Curved Surfaces. arXiv:1601.00324 [cond-mat.soft].
[25] Hussain S, Wivagg CN, Szwedziak P, Wong F, Schaefer K, Izore T, Renner LD, Sun Y, Bisson Filho AW, Walker S, et al. *MreB Filaments Create Rod Shape By Aligning Along Principal Membrane Curvature.* [Internet]. 2017.

# APPENDIX A: GAINING INTUITION FOR THE GROWTH LAW

It is difficult to gain intuition for the terms in Eq. (17). To make our life easier we will neglect terms of order $O(\lambda \epsilon)$, which is a reasonable approximation if $H\lambda \gg \epsilon$. This approximation leaves only the two leading order stress coupled terms, which is sufficient for our current purposes.

The best way to gain intuition is to consider the effect of the various terms on the evolution of special surfaces. Inspired by rod-like *E. coli*, in this paper we focus mainly on elongating cylindrical shapes. However, we will point out how our analysis could be applied to different shapes such as spherical and flat shapes, which are relevant for other interesting growth process as in blooming lilies and rippling leaves [2, 3].

We start with the simplest term, $\alpha_1 g_{ij}$. For very thin surfaces and in the absence of stretching, we can assume that $\bar{g}_{ij} = g_{ij}$. Therefore with time, the metric will evolve as $g_{ij}(t) = e^{\alpha_1 t} g_{ij}(0)$. Therefore the metric is expanding or contracting exponentially regardless of the initial shape. Furthermore, the linear dimensions of the shell grow exponentially at the rate $\alpha_1/2$.

The next term to consider is $\alpha_2 b_{ij}$. This term comes from a simple equation of motion $\partial_t \mathbf{X} = -\alpha_2 \hat{N}$, which is easy to check with the relation $\partial_t g_{ij} = \partial_t \partial_i \mathbf{X} \cdot \partial_j \mathbf{X} + \partial_i \mathbf{X} \cdot \partial_t \partial_j \mathbf{X}$. This evolution is volume minimizing when $\alpha_2$ is positive. In the case of cylindrical growth, this term will tend to shrink the radius. In the absence of growth at the end caps, which appears to be approximately true for *E. coli* [12], the length will not be affected by this growth term. This gives us a way to fix the radius during exponential elongation. Specifically, if we set $\alpha_1 = \alpha_2$ we get an exponentially elongating cylinder with fixed radius.

In the case of a growing sphere, setting $\alpha_1 = \alpha_2$ would just stop the growth. On the other hand, the metric of a flat shell would not be affected by this term at all.

Next we consider the term $\beta_1 (H - H_0) g_{ij}$, where we've subtracted $H_0$ by changing the definition of $\alpha_1$. This terms couples mean curvature to the rate of growth. It depends on the actual shape and not just on the value of the metric. There is no simple interpretation for $\partial_t \mathbf{X}$ in this case. As we will show, this term with $\beta_1 > 0$ is important for the stability of modes with long wavelength in the longitudinal direction and short wavelength in the azimuthal direction.

Note that the $\alpha_2$ and $\beta_1$ terms are dependent on the global orientation of the normal vector $\hat{N}$, which follows from the definition of the curvature tensor. This was not mentioned in Sec. (IV A) as a problem for the invariance of the growth law because we assume the growth process can distinguish between the inside and the outside of the shell. This is not hard to accept in the case of E. *coli* for example. However, for a growing leaf or flower, it might not be possible to distinguish in from out. Therefore, for growing open shells such as leaves, the terms $\alpha_2 b_{ij}$ and $\beta_1 H g_{ij}$ will be absence due to symmetry considerations.

The term $\beta_2 (H - H_0) b_{ij}$ is related to the mean curvature flow. It would result from the motion $\partial_t \mathbf{X} = -\beta_2 (H - H_0) \hat{N}$ and tends to minimize the area when $\beta_2 > 0$. The stabilizing effect of this term is clear. For a cylinder (or a sphere) with radius $a(t)$, we would get $\dot{a} = 0.5 \, a_0^2 \, \beta_2 \, (1/a - 1/a_0)$. The solution to this equation approaches $a_0$ as $t \to \infty$, behaving like $a \sim e^{-\beta_2 t}$.

The last geometric growth term we will consider is related to the well known Ricci flow. Namely $\partial_t g_{ij} = -\gamma_1 \, g_{ij}$. It is a function of the metric only and we do not need to find the corresponding shape to solve this equation. In order to understand the effect of this term, let's switch to a coordinate system such that the metric can be written in the form $g_{ij} = e^\rho \delta_{ij}$, this form will be preserved under evolution since the equation can now be expressed as.

$$\dot{\rho}\, e^\rho = -\frac{\gamma_1}{2} e^\rho \left(-e^\rho \nabla^2 \rho\right) \implies \dot{\rho} = \frac{\gamma_1 e^{-\rho}}{2}\nabla^2 \rho. \quad \text{(A1)}$$

Notice the resemblance of this equation to the diffusion equation. Indeed, in the vicinity of a cylinder we have $\rho \sim 0$, then to leading order this equation becomes exactly the diffusion equation, which tends to wash out the deformations over time, returning the metric back to the constant flat metric.

Finally we consider the two stress coupling terms. The term $\sigma_1\,\epsilon_{ij}$ will tend to make the target metric $\bar{g}_{ij}$ evolve towards $g_{ij}$ when $\sigma_1 > 0$. In other words, it makes the surface comply with the applied forces, as in the case of *E. coli* [9].

The term $\sigma_2\,\epsilon$ couples the areal strain to the growth rate, ignoring the shear strain. For positive $\sigma_2$, and for a given areal strain $\epsilon$, this will make the surface stretch or compress isotropically in a manner proportional to $\epsilon$.

In Sec. (V), we will explore how all these terms interact to generate a linearly stable elongating cylinder. But first lets gain more intuition by looking at various toy models of growth processes.

## APPENDIX B: MICROSCOPIC TOY MODELS

In this section we will consider various toy (or semi-realistic!) models of growth processes. This will give us valuable insight into how the various terms in Eq. (17) might appear and the order of magnitude of their coefficients.

The first model is inspired by the process of swelling polymers films [7, 8]. When polymer films are exposed to a solvent, the solvent molecules will diffuse through the pores in the film and cause swelling of the material. The local rate of swelling can be controlled by different external stimuli such as light and chemical gradients. In the present model we will consider the heterogeneous swelling caused by the curvature of the shells, assuming that only the inner surface is exposed to the solvent.

Since only one side of the shells is exposed then the rate of solvent absorption will depend on the average pore area in the exposed surface. In order to understand the effect of curvature on the exposed pore area, we express the exposed surface $\mathbf{X}_{\text{exp}}$ in terms of the mid-surface of the shell in a manner consistent with the Kirchhoff-Love assumptions. Specifically,

$$\mathbf{X}_{\text{exp}} = \mathbf{X} - \frac{\tau}{2}\,\hat{\mathbf{N}}. \quad \text{(B1)}$$

We can use this relation to relate the area element in the exposed surface $dA_{\text{exp}}$ to the area element in the mid-surface $dA$ using the relation $dA_{\text{exp}} \equiv \sqrt{g_{\text{exp}}}\; du^1 du^2$. Using Eq. (B1), we can relate the two metrics using the formula

$$g_{ij}^{\text{exp}} = g_{ij} + \tau\, b_{ij} + \frac{\tau^2}{4} b_i^\ell\, b_{\ell j}. \quad \text{(B2)}$$

Then, using this relation and the identity $\det(M) = \exp[Tr\{\log(M)\}]$, we can find the relation between the two area elements as

$$\frac{dA_{\text{exp}}}{dA} = 1 + \tau H + \frac{\tau^2}{4} K. \quad \text{(B3)}$$

Finally, allowing for the possibility of strain $\epsilon_{ij}$, we can relate the target and actual mid-surface area elements as

$$dA = \left(1 + \frac{\epsilon}{2} + O\left(\epsilon^2\right)\right)\, d\bar{A}, \text{ where } \epsilon = \bar{g}^{ij}\,\epsilon_{ij} \quad \text{(B4)}$$

In these systems the growth process is isotropic, meaning that only terms of the form $F_1(H,K,\epsilon)\, g_{ij}$ will contribute. Since curvature and strain change the area element of the inner surface by the given geometric factor, we conclude that the average exposed pore area will be affected by the same factor. Finally, assuming that the absorption rate in the absence of curvature and strain is given by $\alpha_1$, we can write the growth law as $\partial_t(d\bar{A})/dA_{\text{exp}} = \alpha_1$.

Putting all of this together we get in the curved case that

$$F_1(H,K) = \alpha_1 \left(1 + \tau H + \frac{\tau^2}{4} K\right)\, \left(1 + \frac{\epsilon}{2}\right). \quad \text{(B5)}$$

Note that the term $H^2$ does not appear in this formula due to cancellations in the calculation of the determinant. In addition, the Ricci flow term is suppressed by an additional power of the thickness. Interestingly, the strain coupling term $\epsilon\, g_{ij}$ in Eq. (17) will have a coefficient $\sigma_2 = \alpha_1/2$.

The terms proportional to the tensors $b_{ij}$ and $\epsilon_{ij}$ are not generated if the growth is isotropic.

Next, we describe toy models where the growth rate of a shell depends on the local concentration of some particle on the surface. This is similar to E. *coli* where the local concentration of the protein MreB affects the growth rate of the cell wall [12]. Here we will describe a simple model of passively diffusing particles on the surface. The heterogeneity results from the dependence of the adhesion energy on the local curvature [22, 23]. Another method for achieving heterogeneity would be active particles moving inside or on the surface of the shell [24]

Fig. (B1) shows a simple diffusing particle composed of two identical orthogonal filaments each with a natural curvature $\bar{\kappa}$ and length $\bar{\ell}$. Assuming that the particles adheres strongly to the surface we can take the realized curvatures of the filaments $(\kappa_{R1}, \kappa_{R2})$ to be determined by the principle curvatures of the surface and the angle $\theta$ between the filaments and the principle directions. Explicitly,

$$\kappa_{R1} = \cos(\theta)^2\, \kappa_1 + \sin(\theta)^2\, \kappa_2, \quad \text{(B6)}$$

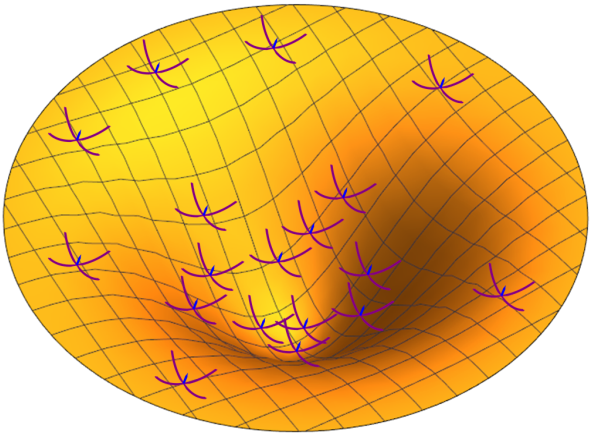

FIG. B1: Shows a "molecule" composed of two perpendicular filaments each with natural curvature $\bar{\kappa}$ and length $\bar{\ell}$ diffusing passively on the surface. The vector shown points to the inside of the closed surface when the filament is attached. The difference in density along the surface is due to biased diffusion based on curvature dependence of adhesion energy

where $\kappa_{1,2}$ are the principle curvatures of the surface. $\kappa_{R2}$ follows a similar expression with $\theta \to \pi/2 - \theta$. If we take the bending energy of each filament to be $E_{bi} = K_b \, \bar{\ell} \, (\kappa_{Ri} - \bar{\kappa})^2$, then the total energy in terms of the Gaussian and mean curvatures will be

$$E_b = K_b \, \bar{\ell} \, \bar{\kappa}^2 \times \left( 2 - 4\frac{H}{\bar{\kappa}} + \frac{3H^2 - K}{\bar{\kappa}^2} + \frac{H^2 - K}{\bar{\kappa}^2} \cos(4\theta) \right). \quad \text{(B7)}$$

We can easily see that this energy is minimized when $\theta = \pi/4$. The steady state concentration of randomly diffusing particles will be given by a Boltzmann factor $\rho \propto \exp(-\beta \, E_b)$. We also assume, as we have throughout the paper, that the curvatures of the surface are small compared to the natural curvature of the filament $H \ll \bar{\kappa}$. Assuming a growth rate proportional to concentration $\partial_t(dA)/dA = C \, \rho$, we get finally that

$$\partial_t \bar{g}_{ij} = \left( \alpha_1 + \beta_1 \, a_0 H + \delta_1 \, \frac{H^2}{\bar{\kappa}^2} + O\left(\frac{H}{\bar{\kappa}}\right)^3 \right) g_{ij}, \quad \text{(B8)}$$

were $\alpha_1 \equiv C \, \rho_0$ with $\rho_0$ being the concentration of the particles when the surface is flat and C being a constant relating the growth rate to the concentration. We also defined

$$\beta_1 \equiv \frac{4 \, \beta \, K_b \, \bar{\ell} \, \bar{\kappa} \, \alpha_1}{a_0} \approx 1.3 \, \alpha_1, \quad \text{(B9)}$$

$$\delta_1 \equiv \frac{\beta_1 \, (\bar{\kappa} \, a_0 \, \beta_1 - \alpha_1)}{\bar{\kappa} \, a_0 \, \alpha_1}, \quad \text{(B10)}$$

where the parameter $\beta_1$ was estimated at room temperature, for an MreB-like filament with 10 monomers ($\bar{\ell} \approx 50$nm) and following Ref. [25], $K_b \bar{\kappa}^2 \approx 8.2 \times 10^{-13}$. Finally, we assumed that $a_0 \bar{\kappa} = 0.1$.

It is intriguing – keep in mind that the estimate could be wrong by a couple of orders of magnitude in either direction – that this number came out to be of order 1. Although we don't consider strain coupling in this model, we expect it to also behave as $\sigma/\alpha_1 \sim O(1)$, just as it did in the absorption model above. The fact that $\sigma$ is on the same order as $\alpha_1$ was a natural consequence of the growth. On the other hand, $\beta_1 \sim \alpha_1$ depends on the actual value of temperature and rigidity of MreB.

As mentioned in appendix (A), the term $H \, g_{ij}$ is dependent on the definition of the normal to the surface. It appears in Eq. (B8) because we assumed the filament attaches to the inner surface with the arrow pointing opposite to $\hat{N}$. If we relax this assumption or consider an energy like $E_b \sim (\kappa_R^2 - \bar{\kappa}^2)^2$, this term disappears and the leading order terms will be $H^2 g_{ij}$ and $K g_{ij}$.

## APPENDIX C: SCALING BEHAVIOR FOR SMALL WAVELENGTHS

In this section, we will study more closely the growth of modes with small wavelengths, namely $q_P \to \infty$. We will gain insight by contrasting the finite and zero thickness cases, starting with the latter.

As can be seen from Eq. (28), modes were $q_P \to \infty$ can be stabilized by requiring $\Gamma_2 > 0$. This term ultimately comes from the growth terms $H g_{ij}$, $H b_{ij}$ and $K g_{ij}$ in Eq. (18). Furthermore, we can easily show that as $q_P \to \infty$, these terms scale as

$$H \, g_{zz} \sim H \, b_{zz} \sim K \, g_{zz} \sim q_P^2 \, \rho_{mq}. \quad \text{(C1)}$$

We can also easily see, from the isometric solutions given in Eq. (25), that $\rho_{mq} \sim G_{\phi\phi}$. This, together with Eq. (C1) leads to the stabilizing term $q_P^2 \, \Gamma_2$ in Eq. (28).

Now we can understand qualitatively how finite thickness would change this result. Bending energy suppresses deformations that have wavelengths comparable to thickness, specifically, we get $\rho_{mq} \sim G_{\phi\phi}/q_P^4$. Therefore the stabilizing term proportional to $\Gamma_2$ would disappear as $q_P \to \infty$. Next, we examine this case a little more concretely.

First, we minimize the energy with a given metric deformation $G_{ij}$, and solve for the displacements $\rho_{mq}, h_{mq}$ and $\psi_{mq}$. we get, for example, that

$$\rho_{mq} \sim O\left(\frac{G_{\phi\phi}}{q_P^4}\right) + O\left(\frac{G_{z\phi}}{q_P^5}\right) + O\left(\frac{G_{zz}}{q_P^6}\right). \quad \text{(C2)}$$

We then plug these solutions back into the growth law and get, for example, that $\partial_t G_{\phi\phi} \propto -\Gamma_2 \, G_{\phi\phi}/q_P^2$. Thus we see that $G_{\phi\phi}$ can still be stabilized if $\Gamma_2 > 0$ and in what follows, we set $G_{\phi\phi} \to 0$. The other two equations give

$$\partial_t G_{zz} \sim 2 \, R_0 \, G_{zz} \quad \text{and} \quad \partial_t G_{z\phi} \sim 2 \, R \, \frac{G_{zz}}{q_P}. \quad \text{(C3)}$$

Finally, combining Eqs. (C3) and (C1) we discover that

$$\frac{\partial_t \rho_{mq}}{\rho_{mq}} = 4R_0 + O\left(\frac{1}{q_P}\right), \tag{C4}$$

which validates the result obtained in Eq. (30).